\documentstyle [12pt] {article}

\catcode`@=12
\newcommand{\be}{\begin{equation}}
\newcommand{\ee}{\end{equation}}
\newcommand{\ber}{\begin{eqnarray}}
\newcommand{\eer}{\end{eqnarray}}
\newcommand{\bers}{\begin{eqnarray*}}
\newcommand{\eers}{\end{eqnarray*}}
\newcommand{\bt}{\begin{itemize}}
\newcommand{\et}{\end{itemize}}
\begin{document}
\vspace{0.5in}
\oddsidemargin -.375in
\newcount\sectionnumber
\sectionnumber=0
\def\bra#1{\left\langle #1\right|}
\def\ket#1{\left| #1\right\rangle}
\def\be{\begin{equation}}
\def\ee{\end{equation}}
\thispagestyle{empty}
\begin{flushright} UdeM-GPP-TH-01-81 \\February 2001\
\end{flushright}
\vspace {.5in}
\begin{center}
{\Large\bf Weak decays to final states with
Radial Excitation Admixtures  \\}
\vspace{.5in}
{{\bf Alakabha Datta{\footnote{email: datta@lps.umontreal.ca}} 
${}^{a}$}, {\bf Harry J. Lipkin {\footnote{email: 
harry.lipkin@weizmann.ac.il}} 
${}^{b}$} and
{\bf Patrick. J. O'Donnell {\footnote{email:
pat@medb.physics.utoronto.ca}}${}^{c}$} 
 \\}
\vspace{.1in}
${}^{a}$
{\it Laboratoire Ren\'e J.-A. L\'evesque, Universit\'e de
Montr\'eal,} \\
{\it C.P. 6128, succ.\ centre-ville, Montr\'eal, QC, Canada H3C 3J7} \\
${}^{b)}$ {\it
Department of Particle Physics,\\
Weizmann Institute,\\
Rehovot 76100, Israel \\and\\
School of Physics and Astronomy, \\
Tel-Aviv University,\\
Tel-Aviv 69978, Israel \\
${}^{c}$ {\it Department of Physics and Astronomy,\\
University of Toronto, Toronto, Canada.}\\
}
\end{center}

\begin{abstract}
We consider the weak decays of a B meson to final states that are mixtures 
of S-wave radially excited components. We  consider non leptonic 
decays of the type $B \to \rho' \pi/B \to \rho \pi$, 
$B \to \omega' \pi/B \to \omega \pi$ and
$B \to \phi' \pi/B \to \phi \pi$ 
where $\rho'$, $\omega'$ and $\phi'$ are higher $\rho$, $\omega$ 
and $\phi$ resonances. We find
 such decays to have  larger or similar branching ratios
compared
to decays where
the final state $\rho$, $\omega$ and $\phi$ are in the ground state.
We also study the effect of radial mixing in the vector and 
the pseudoscalar systems generated from hyperfine interaction 
and the annihilation
term. We find the effects of radial mixing to be small and generally
 negligible for all practical purposes in the vector system.
However, in  the $\eta-\eta^{\prime}$ system the effects 
of radial mixing are appreciable and
seriously affect decay branching ratios for
 $B \to \eta(\eta')K(K^*)$. In particular we 
find that nonstandard $\eta(\eta')$ mixing can resolve  the puzzles
in $B \to \eta(\eta')K$ decays.
 
\end{abstract}

\newpage \pagestyle{plain}

\section{Introduction}

Nonleptonic $B$ decays play a very important role
in the study of CP violation. It is expected that these studies will test the standard model(SM) picture of CP violation or provide hints for new physics.
Most studies of two body nonleptonic $B$ decays have concentrated on
processes of the type $B \to M_1 M_2$ where both $M_1$ and $M_2$ 
are mesons in the ground state configuration. Here we want to look at
nonleptonic B decays to final states where one of the final state meson
contain admixtures of radially excited components. 
We expect such decays to have  larger or similar branching ratios
compared
to decays where
the final state contains the same meson in the ground state. There is an 
easy explanation for such a statement.
For simplicity let us consider a simple model in which
$B \rightarrow \pi M$ and M is a simple flavor eigenstate with no flavor
mixing beyond isospin;
e.g. $\rho$, $\omega$ or $\phi$ and we are interested in comparing 
the branching 
ratios
for the final states containing the ground state configuration of meson $M$ 
and
its radial excitation. 
Some possible examples are:
\begin{equation}
BR({\bar{B}^0} \rightarrow \pi^- \rho^{+ \prime} )/
BR({\bar{B}^0} \rightarrow \pi^- \rho^{+} )
\end{equation}
\begin{equation}
BR(B^- \rightarrow \pi^- \omega^{\prime} )/BR(B^- \rightarrow \pi^- \omega)
\end{equation}
\begin{equation}
BR(B_s \rightarrow \pi^0 \phi^{\prime} )/BR(B_s \rightarrow \pi^0 \phi)
\end{equation}
where  $\rho^{+ \prime}$, $\omega^{\prime}$ and $\phi^{\prime}$ are radially excited states.

We assume an extreme factorization approximation in 
which
the $b$ quark decays into a pion and a $u$-quark and we neglect the relative
 Fermi
momentum of the initial $b$ and the spectator $\bar q$. 
The quark transition  for the processes (1) and (2) is then
\begin{equation}
b \rightarrow \pi^-(\vec p) u(-\vec p)
\end{equation}
where $\vec p$ denotes the final momentum of the $\pi^-$.
For the process (3) the quark transition is essentially similar to 
the one above
\begin{equation}
b \rightarrow \pi^0(\vec p) s(-\vec p)
\end{equation}
where now $\vec p$ denotes the final momentum of the $\pi^0$.

Concentrating on the processes (1) and (2)
the transition matrix for the full decay has the form
\begin{equation}
\bra {\pi^-(\vec p) M(-\vec p)}T\ket {B} =
\bra {M(-\vec p)}F \ket{u(-\vec p) \bar q(0)}\cdot
\bra {\pi^-(\vec p) u(-\vec p) ) }W \ket {b}
\end{equation}
where $T$ denotes the transition matrix for the hadronic decay which factors
into a weak matrix element at the quark level denoted by $W$ and a 
fragmentation
matrix element denoted by $F$ describing the transition of a quark with 
momentum
$- \vec p$ and an antiquark with zero momentum to make a meson with momentum
$- \vec p$.

It is immediately clear that if the final momentum $ \vec p$ is large, the
fragmentation matrix element will depend upon the high momentum tail of the
meson wave function. This might tend to favor radial excitations over ground
states, since the radial excitations are expected to have higher kinetic
energies.
We now note that the harmonic oscillator wave functions commonly used 
in
hadron spectroscopy have a Gaussian tail for their high momentum components 
and
 this can suppress the fragmentation matrix element
$\bra {M(-\vec p)}F \ket{u(-\vec p) \bar q(0)}$ in comparison with
wave functions from a different confining potential. Hence the branching ratio
to a final state which is radially excited will be sensitive to the choice of the confining potential.

So far we have assumed the physical states to be pure radial excitations.
However additional interactions can mix the various radial excited components.
For instance hyperfine interactions can mix radial 
excitations with the same flavor structure and so in general in
the
$\rho$, $\omega$ and $\phi$ system the various physical states will 
be admixtures of radial excitations \cite{Cohen:1979ge,Frank:1984bj}. 
Flavor mixing in the vector system 
is known to be small but is important
in the pseudoscalar sector. Here the mixing 
in the $\eta(\eta')$ system receives an additional significant contribution 
from the annihilation diagram that leads to flavor mixing
of the strange and non strange parts of the $\eta(\eta')$ wavefunction. The mixing
 in the pseudoscalar sector, therefore, is different 
 from the ideal mixing found in the $\omega-\phi$ system. It 
is also possible
for the annihilation term to mix states that are radial excitations 
allowing the
the $\eta(\eta')$ wavefunction to contain radially excited 
components 
. Such non standard $\eta(\eta')$ mixing can have important
implication
for the non leptonic decays $B \to \eta(\eta')K(K^*)$. 

In the transitions chosen in (1),(2) and (3)
the radially excited meson must include the
spectator quark and therefore must depend upon the fragmentation matrix 
element
$\bra {M(-\vec p)}F \ket{u(-\vec p) \bar q(0)}$. For the case of
$B \to \eta(\eta')K$ decays,
factorization results in the kaon leaving the weak
vertex with its full momentum and the remaining quark carries the full 
momentum
of the final meson. There is therefore a form factor in which there is a 
large
internal momentum transfer needed to hadronize this quark with the spectator
antiquark. This might favor the radial excitation if it has a much higher 
mean
internal momentum. However things are more complicated here as
the $\eta(\eta')$
can also be produced by an $q\bar q$ pair
in a penguin diagram without
containing the spectator quark. One possibility is when  the $\bar{s}$ 
quark in 
the QCD penguin combines with the $s$ quark from the $ b \to s$ transition 
to form the $\eta(\eta')$. Another possibility is when a $\bar{q}$ and 
$q$ pair
(where $ q=u,d,s$), 
 appearing in the same current in the effective Hamiltonian, hadronizes to
 the 
$\eta(\eta')$. In the diagrammatic language this is often represented as
a ``gluon'' splitting into a $q \bar q$ pair which then hadronizes into
a $\eta(\eta')$. This term is usually called OZI suppressed 
\cite{Okubo:1963fa,Zweig}
 as in most 
decays the contribution of this term is indeed suppressed with respect
 to other terms in the decay amplitude. This may not be the case
in the $B \to \eta(\eta')K$ decays where the OZI term can be of comparable size as
other terms in the decay amplitude and in particular 
 we show that 
$B \to \eta^{\prime} K^*$ can have significant contribution from the
OZI suppressed term.

This paper is organized in the following manner:
In the next section we study the mass mixing in the vector meson sector, 
involving
the $\rho$, $\omega$ and $\phi$, and the pseudoscalar sector involving the
$\eta(\eta')$. In  section 3 we present a 
general treatment of nonleptonic $B$ decays 
using the effective Hamiltonian and the  factorization assumption. We 
then show how this approach is related to
  the diagrammatic approach of studying nonleptonic $B$ decays.
In  section 4
we study nonleptonic 
decays of $B$ to final states involving 
higher resonant $\rho$, $\omega$ and $\phi$ 
states. This is followed by  section 5 where we make predictions for
 $B \to \eta(\eta^{\prime}) K$ and 
 $B \to \eta(\eta^{\prime}) K^*$  
 and comment on the relevance 
of the OZI suppressed term in the calculation of these decays. 
Finally we present our conclusions. 
\section{Mass mixing in the vector meson and pseudoscalar sector}
We start with the mixing for the $\rho$ system. 
To obtain the eigenstates and eigenvalues we 
diagonalize
the mass matrix which has the form
\ber
< q_a'\bar{q}_b',n'|M|q_a \bar{q}_b,n> & = & 
\delta_{aa'} \delta_{bb'} \delta_{nn'}
(m_a +m_b +E_n) +\delta_{aa'}\delta_{bb'}\frac{B}{m_am_b}
{ \vec{s}_a \cdot \vec{s}_b} \psi_n(0)\psi_{n'}(0)\
\label{mm}
\eer
where $\vec{s}_{a,b}$ and $m_{a,b}$ are the quark spin operators and masses.
Here $n=0,1,2$ and the basis states are chosen as 
$|N>=| u \bar{u} + d \bar{d}>/\sqrt{2}$ and $S=|s \bar{s}>$.
In the above equation $E_n$ is the binding energy of the $n^{th}$ radially excited state and $B$ is the strength of the hyperfine interaction. Note we 
are only presenting results for the neutral mesons. A similar 
treatment also can be used for the charged mesons. We will use a very 
simple model for confinement in our calculations as we do not intend to present
a detailed study of light meson spectroscopy. Our aim, as already stated in the previous section, is to 
study non leptonic decays of the $B$ meson to radially excited 
light meson states as well as to study the effects of radial mixings in
 the non leptonic decays $B \to \eta(\eta')K(K^*)$. We believe 
the conclusions reached on the basis of our calculations are 
likely to hold true in a more 
detailed model of confinement. 

To begin with, we use the same harmonic 
confining potential as well as the other parameters
used in 
Ref\cite{Frank:1984bj} 
to obtain the eigenstates and
eigenvalues for the mass matrix in Eqn.~\ref{mm}. The various 
parameters used in the calculation are $m_u=m_d=$ 0.350 GeV, $m_s=$ 0.503 GeV, 
the angular frequency,
$\omega= $0.365 GeV and $b=B/m_u^2=$0.09. 
  
We obtain for the eigenvalues and eigenstates in the $\rho$ system
\ber
|\rho(0.768)> &= & 0.990|N>_0 +0.124|N>_1-0.066|N>_2 \nonumber\\
|\rho(1.545)> &= & 0.108|N>_0 -0.973|N>_1+0.204|N>_2 \nonumber\\
|\rho(2.370)> &= & 0.089|N>_0 -0.195|N>_1+0.977|N>_2 \
\eer
To see how this result
 changes with a different confining potential we use a power law
potential $V(r)= \lambda r^n$ \cite{Quigg:1979vr} . We will use a linear and a quartic 
confining potential 
and compare the spectrum with that obtained 
with a 
harmonic oscillator potential. To fix the coefficient $\lambda$
we require that the energy eigenvalues of the Schr\"odinger equation are similar 
in the least
square sense with the energy eigenvalues used in
Ref\cite{Frank:1984bj}. So for example, for the linear potential, we 
demand that
$$F=\sum_n (E_n(harmonic)-E_n(linear))^2$$ is a minimum. This fixes the 
constant
$\lambda$ in $V(r)=\lambda r$ and we obtain
\ber
|\rho(0.775)> & =& 0.992|N>_0 +0.112|N>_1-0.053|N>_2 \nonumber\\
|\rho(1.515)> &= & 0.104|N>_0 -0.986|N>_1+0.130|N>_2 \nonumber\\
|\rho(2.260)> & =&0.066|N>_0 -0.122|N>_1+0.99|N>_2 \
\eer
We follow the same procedure 
for the quartic potential and obtain
\ber
|\rho(0.759)> & =& 0.988|N>_0 +0.129|N>_1-0.077|N>_2 \nonumber\\
|\rho(1.567)> &= & 0.103|N>_0 -0.955|N>_1+0.278|N>_2 \nonumber\\
|\rho(2.550)> &= & 0.11|N>_0 -0.267|N>_1+0.957|N>_2 \
\eer

We observe that the mass eigenstates and 
eigenvalues are not very sensitive to the confining potential and the 
radial mixing effects are  small.

We next turn to mixing in the $\eta-\eta'$ system.
In the traditional picture the $\eta$ and $\eta^\prime$ mesons are mixtures 
of
singlet and octet states
$\eta_1$ and $\eta_8$ of $SU(3)$.
\ber
\left[ \begin{array}{c} \eta \\ \eta' \end{array} \right]
&=& \left[
\begin{array}{cc}
\cos \theta & - \sin \theta \\
\sin \theta & \cos \theta
\end{array}
\right]
\left[
\begin{array}{c}
\eta_8 \\ \eta_1
\end{array}
\right] \\
\eta_8 &=& \frac{1}{\sqrt{6}} \left[ u \bar{u} +
d \bar{d} -2 s \bar{s} \right] \\
\eta_1 &=& \frac{1}{\sqrt{3}} \left[ u \bar{u} +
d \bar{d} + s \bar{s} \right]
\eer
where the mixing angle $\theta$ lies between $-10^0$ and $-20^0$\cite{rpp2000}.

To obtain the eigenstates and eigenvalues in the $\eta-\eta'$ system we 
diagonalize
the mass matrix
\ber
< q_a'\bar{q}_b',n'|M|q_a \bar{q}_b,n> & = & 
\delta_{aa'} \delta_{bb'} \delta_{nn'}
(m_a +m_b +E_n) +\delta_{aa'}\delta_{bb'}\frac{B}{m_am_b} 
{\vec{s}_a \cdot \vec{s}_b} \psi_n(0)\psi_{n'}(0)
\nonumber\\
& + & \delta_{ab} \delta_{a'b'} \frac{A}{m_am_b} \psi_n(0)\psi_{n'}(0)\
\label{massmatrix}
\eer
This has a similar structure as the $\rho$ system but now we have the
additional annihilation contribution with strength $A$ that causes 
flavor mixing. In our calculations we  use the
  phase convention in Ref\cite{Frank:1984bj} where the wavefunctions 
at the origin in 
configuration space, which enter
 in the hyperfine and annihilation terms in the mass matrix, 
are positive(negative) for the even(odd) radial excitations.

We try to fit the values of $A$ and $B$ to the measured masses. The mass matrix 
is $6 \times 6$ matrix which we diagonalize to make predictions for 6 masses and mixings.
However, for the sake of brevity we will only give the predictions 
for $\eta$ and $\eta'$ masses and wavefunctions.
Several 
solutions that give acceptable values of the masses can be obtained. We choose
 solutions for the linear, quadratic and quartic confining potentials
that make similar predictions for the $\eta(\eta^{\prime})$ masses:

For the linear potential we obtain with $B=0.065m_u^2$ and $A=0.045m_u^2$
\ber
\ket{\eta(0.544)} &  = &0.961|N>_0 -0.198|N>_1+0.108|N>_2 \nonumber\\  
& - & 0.150|S>_0 
 +  0.050|S>_1
-0.032|S>_2\nonumber\\
\ket{\eta'(0.924)} &= &0.170|N>_0 +0.0385|N>_1-0.0157|N>_2\nonumber\\ 
& + & 0.974|S>_0 
-  0.126|S>_1 
 +  0.0486|S>_2\
\eer

For the harmonic potential we obtain with $B=0.065m_u^2$ and $A=0.065m_u^2$
\ber
\ket{\eta(0.547)} & =& 0.913|N>_0 -0.252|N>_1+0.154|N>_2 \nonumber\\ 
& - & 0.249|S>_0  
 +  0.107|S>_1
-0.076|S>_2\nonumber\\
\ket{\eta'(0.931)} & =& 0.316|N>_0 +0.109|N>_1-0.049|N>_2\nonumber\\ 
& + & 0.925|S>_0  
- 0.148|S>_1 
 +  0.088|S>_2\
\eer
Our results for the harmonic potential is similar to the results 
obtained in  Ref\cite{Frank:1984bj} where
 a slightly different mass mixing matrix than the one used here has 
been used to obtain the $\eta-\eta^{\prime}$ mixing.

Finally, for the quartic potential we obtain with 
$B=0.065m_u^2$ and $A=0.11m_u^2$
\ber
|\eta(0.547)> &= & 0.764|N>_0 -0.287|N>_1+0.198|N>_2\nonumber\\ 
& - & 0.441|S>_0  
 +  0.248|S>_1
-0.196|S>_2\nonumber\\
|\eta'(0.940)> &= & 0.623|N>_0 +0.350|N>_1-0.177|N>_2 \nonumber\\  
&+ & 0.658|S>_0 
 -  0.134|S>_1 
 +  0.087|S>_2\
\label{quartic}
\eer

It is clear that the eigenstates of $\eta(\eta')$ 
system are sensitive to the 
confining potential and there can be substantial radial mixing 
 which can then affect the predictions
for the decays $B \to \eta(\eta') K(K^*)$ . We  note that as
we move from the linear to the quartic potential the $\eta(\eta')$  
mixing deviates
more significantly from the ideal mixing case. This may be understood from the fact that
the fit to annihilation term, $A$, is the largest for the quartic potential
which leads to the largest deviation from the ideal mixing case. A standard 
$\eta-\eta^{\prime}$ 
mixing often used in the literature is given by
\cite{Bramon:1974ky,Lipkin:1997ad}
\ber
\ket{\eta^{\prime}}_{std} & = & \frac{1}{\sqrt{6}} \left[ u \bar{u} +
d \bar{d} +2 s \bar{s} \right] \nonumber\\
\ket{\eta}_{std} &=& \frac{1}{\sqrt{3}} \left[ u \bar{u} +
d \bar{d} - s \bar{s} \right]
\eer
We can then write the 
$\eta-\eta^{\prime}$  
states obtained with the various confining potential and keeping 
only the ground states, in terms of the the states defined above. For 
the linear potential we find
\ber
\ket{\eta^{\prime}} & = & 0.89\ket{\eta^{\prime}}_{std} -0.43 \ket{\eta}_{std}
\nonumber\\
\ket{\eta} & = & 0.87\ket{\eta}_{std} +0.43 \ket{\eta^{\prime}}_{std}\
\eer
For the harmonic potential
\ber
\ket{\eta^{\prime}} & = & 0.94\ket{\eta^{\prime}}_{std} -0.27 \ket{\eta}_{std}
\nonumber\\
\ket{\eta} & = & 0.89\ket{\eta}_{std} +0.32 \ket{\eta^{\prime}}_{std}\
\eer
 and finally for the quartic potential one finds
\ber
\ket{\eta^{\prime}} & = & 0.90\ket{\eta^{\prime}}_{std} +0.13 \ket{\eta}_{std}
\nonumber\\
\ket{\eta} & = & 0.88\ket{\eta}_{std} +0.08 \ket{\eta^{\prime}}_{std}\
\eer
This shows that all three confining potentials give mixings
for the $\eta-\eta^{\prime}$   that have substantial overlap with the
standard mixing but the mixing with the quartic potential is closest to 
the standard mixing in the sense that
here one has the smallest
component of the $\ket{\eta}_{std}(\ket{\eta^{\prime}}_{std}$ in 
$\ket{\eta^{\prime}}(\ket{\eta})$.

We now turn to the $\omega-\phi$ system. As in the
the $\eta-\eta'$ system we 
diagonalize
the mass matrix in Eqn.~\ref{massmatrix}.
 We use the same value for 
the hyperfine interaction as used for the $\rho$ system. Again, for the 
sake of brevity, we only
 give the wavefunctions for the ground and the first excited states.

For the linear potential we obtain with $B=0.09m_u^2$ and $A=0.005m_u^2$
\ber
\ket{\omega(0.782)} &  = &0.991|N>_0 +0.123|N>_1-0.058|N>_2 \nonumber\\ 
& - & 0.014|S>_0  
 +  0.004|S>_1
-0.002|S>_2\nonumber\\
\ket{\phi(1.05)} &= & 0.012|N>_0 +0.011|N>_1-0.004|N>_2 \nonumber\\  
& + & 0.997|S>_0 
 +  0.071|S>_1 
 -  0.034|S>_2\nonumber\\
\ket{\omega(1.52)} &  = &-0.113|N>_0 +0.982|N>_1+0.144|N>_2 \nonumber\\  
& - & 0.006|S>_0  
 -  0.034|S>_1
+0.004|S>_2\nonumber\\
\ket{\phi(1.66)} &= &0.007|N>_0 -0.030|N>_1-0.014|N>_2 \nonumber\\  
& + & 0.068|S>_0
-  0.994|S>_1 
 -  0.077|S>_2\
\eer

For the harmonic potential we obtain with $B=0.09m_u^2$ and $A=0.015m_u^2$
\ber
\ket{\omega(0.783)} &  = &0.984|N>_0 +0.154|N>_1-0.081|N>_2 \nonumber\\  
&-& 0.033|S>_0  
 +  0.011|S>_1
-0.007|S>_2\nonumber\\
\ket{\phi(1.05)} &= &0.026|N>_0 +0.029|N>_1-0.011|N>_2 \nonumber\\  
&+& 0.994|S>_0 
+  0.089|S>_1 
 -  0.048|S>_2\nonumber\\
\ket{\omega(1.57)} &  = &-0.126|N>_0 +0.948|N>_1+0.256|N>_2 \nonumber\\  
& - & 0.008|S>_0  
 -  0.139|S>_1
+0.010|S>_2\nonumber\\
\ket{\phi(1.68)} &= &0.025|N>_0 -0.118|N>_1-0.07|N>_2 \nonumber\\  
&+& 0.082|S>_0 
-  0.976|S>_1 
 -  0.143|S>_2\
\eer

For the quartic potential we obtain with $B=0.09m_u^2$ and $A=0.023m_u^2$
\ber
\ket{\omega(0.782)} &  = &0.980|N>_0 +0.163|N>_1-0.096|N>_2\nonumber\\  
& - & 0.049|S>_0  
 +  0.012|S>_1
-0.009|S>_2\nonumber\\
\ket{\phi(1.05)} &= &0.041|N>_0 +0.034|N>_1-0.015|N>_2 \nonumber\\  
& +& 0.991|S>_0 
+ 0.100|S>_1 
 -  0.060|S>_2\nonumber\\
\ket{\omega(1.58)} &  = &-0.122|N>_0 +0.932|N>_1+0.322|N>_2 \nonumber\\  
& - & 0.010|S>_0  
 -  0.113|S>_1
+0.006|S>_2\nonumber\\
\ket{\phi(1.7)} &= &0.022|N>_0 -0.089|N>_1-0.067|N>_2 \nonumber\\  
&+& 0.086|S>_0
-  0.968|S>_1 
 -  0.207|S>_2
\eer

As in the $ \rho$ system we find the mixing to be insensitive to the 
confining potential and we also find the effects of radial mixing 
to be small. We also find, as expected,
 a smaller value for the
annihilation term in the fits to the masses as compared to 
the pseudoscalar system.

\section{ Effective Hamiltonian, Factorization and the Diagrammatic approach}

In the Standard Model (SM)
the amplitudes for hadronic $B$ decays 
are generated by the following effective
Hamiltonian \cite{Reina}:
\begin{eqnarray}
H_{eff}^q &=& {G_F \over \protect \sqrt{2}}
[V_{fb}V^*_{fq}(c_1O_{1f}^q + c_2 O_{2f}^q) -
\sum_{i=3}^{10}(V_{ub}V^*_{uq} c_i^u
+V_{cb}V^*_{cq} c_i^c +V_{tb}V^*_{tq} c_i^t) O_i^q] +H.C.\;
\end{eqnarray}
where the
superscript $u,\;c,\;t$ indicates the internal quark, $f$ can be $u$ or
$c$ quark and $q$ can be either a $d$ or a $s$ quark depending on
whether the decay is a $\Delta S = 0$
or $\Delta S = -1$ process.
The operators $O_i^q$ are defined as
\begin{eqnarray}
O_{1f}^q &=& \bar q_\alpha \gamma_\mu Lf_\beta\bar
f_\beta\gamma^\mu Lb_\alpha\;\;\;\;\;\;\;O_{2f}^q =\bar q
\gamma_\mu L f\bar
f\gamma^\mu L b\;\nonumber\\
O_{3,5}^q &=&\bar q \gamma_\mu L b
\bar q' \gamma_\mu L(R) q'\;\;\;\;\;\;\;\;O_{4,6}^q = \bar q_\alpha
\gamma_\mu Lb_\beta
\bar q'_\beta \gamma_\mu L(R) q'_\alpha\;\\
O_{7,9}^q &=& {3\over 2}\bar q \gamma_\mu L b e_{q'}\bar q'
\gamma^\mu R(L)q'\;\;O_{8,10}^q = {3\over 2}\bar q_\alpha
\gamma_\mu L b_\beta
e_{q'}\bar q'_\beta \gamma_\mu R(L) q'_\alpha\;\nonumber
\end{eqnarray}
where $R(L) = 1 \pm \gamma_5$,
and $q'$ is summed over u, d, and s. $O_1$ and $O_2$ are the tree
level and QCD corrected operators. $O_{3-6}$ are the strong gluon induced
penguin operators, and operators
$O_{7-10}$ are due to $\gamma$ and Z exchange (electroweak penguins),
and ``box'' diagrams at loop level. The Wilson coefficients
$c_i^f$ are defined at the scale $\mu \approx m_b$
and have been evaluated to next-to-leading order in QCD.
The $c^t_i$ are the regularization scheme
independent values obtained in Ref. \cite{FSHe}.
We give the non-zero $c_i^f$
below for $m_t = 176$ GeV, $\alpha_s(m_Z) = 0.117$,
and $\mu = m_b = 5$ GeV,
\begin{eqnarray}
c_1 &=& -0.307\;,\;\; c_2 = 1.147\;,\;\;
c^t_3 =0.017\;,\;\; c^t_4 =-0.037\;,\;\;
c^t_5 =0.010\;,
c^t_6 =-0.045\;,\nonumber\\
c^t_7 &=&-1.24\times 10^{-5}\;,\;\; c_8^t = 3.77\times 10^{-4}\;,\;\;
c_9^t =-0.010\;,\;\; c_{10}^t =2.06\times 10^{-3}\;, \nonumber\\
c_{3,5}^{u,c} &=& -c_{4,6}^{u,c}/N_c = P^{u,c}_s/N_c\;,\;\;
c_{7,9}^{u,c} = P^{u,c}_e\;,\;\; c_{8,10}^{u,c} = 0
\label{wc}
\end{eqnarray}
where $N_c$ is the number of colors.
The leading contributions to $P^i_{s,e}$ are given by:
$P^i_s = ({\frac{\alpha_s}{8\pi}}) c_2 ({\frac{10}{9}} +G(m_i,\mu,q^2))$ and
$P^i_e = ({\frac{\alpha_{em}}{9\pi}})
(N_c c_1+ c_2) ({\frac{10}{9}} + G(m_i,\mu,q^2))$.
The function
$G(m,\mu,q^2)$ is given by
\begin{eqnarray}
G(m,\mu,q^2) = 4\int^1_0 x(1-x) \mbox{ln}{m^2-x(1-x)q^2\over
\mu^2} ~\mbox{d}x \;
\end{eqnarray}
All the above coefficients are obtained up to one loop order in electroweak
interactions. The momentum $q$ is the momentum carried by the virtual gluon 
in
the penguin diagram.
When $q^2 > 4m^2$, $G(m,\mu,q^2)$ develops an imaginary part.
In our calculation, we 
use, for the current quark masses at the scale $ \mu \sim m_b$,
 $m_u = 5$ MeV, $m_d = 7$ MeV, $m_s = 100$ MeV, $m_c = 1.35$ GeV
\cite{rpp2000,lg} and use $q^2=m_b^2/2$.

 The structure of the effective Hamiltonian allows us to write the amplitude
for $B \to M_1 M_2$ as
\ber
A & =& {G_F \over \protect \sqrt{2}}
[V_{fb}V^*_{fq}T -
(V_{ub}V^*_{uq}P_u
+V_{cb}V^*_{cq}P_c +V_{tb}V^*_{tq}P_t] \
\eer
Now we can write the tree amplitude $T$ as
\ber
T & = & \bra{M_1,M_2}c_1O_{1f}+c_2O_{2f}\ket{B} \nonumber\\
T & = & c_2\bra{M_1,M_2}O_{2f}-(1/N_1)O_{1f}\ket{B} \
\eer
where from Eqn.~\ref{wc} $N_1=-c_2/c_1=1.147/0.307=3.73$.
In the factorization assumption there are two contributions to the tree
matrix element, $T$. To be specific consider the decay 
$B^- \to \rho^0(\omega) \pi^-$. In this case there can be 
two contributions to $T$, given in the factorization assumption by
\ber
T & =& T_1 +T_2 \nonumber\\
T_1 & = & c_2<\pi^-|{\bar{d}}\gamma^{\mu}Lb|B^->
<\rho^0(\omega)|{\bar{u}}\gamma_{\mu}Lu|0>[\frac{1}{N_c}-\frac{1}{N_1}] 
\nonumber\\
T_2 & = & c_2<\rho^0(\omega)|{\bar{u}}\gamma^{\mu}Lb|B^->
<\pi^-|{\bar{d}}\gamma_{\mu}Lu|0>[1-\frac{1}{N_cN_1}] \
\label{tree}
\eer
In the diagrammatic language the first term, $T_2$, corresponds 
to a $b$ quark
transition to a $u$ quark and a $W^-$ which turns into a $\pi^-$. 
The $u$ quark
then combines with the spectator quark to form 
the $\rho^0(\omega)$ particle. In the term $T_1$, the $u$ quark 
from the $b \to u$ transition combines with the ${\bar{u}}$ quark 
from the $W^-$
to form the $\rho^0(\omega)$ particle while the $d$ quark from the $W^-$
combines with the spectator quark to form the $\pi^-$. This is the 
color suppressed diagram and from the expression above we see that there is an 
additional suppression coming from the Wilson's coefficients and so, effectively
 $T_1$ is suppressed by a factor of $(1/N_c -1/N_1) \sim 1/15$ 
relative to $T_2$.

We now turn to the penguin contribution, and for simplicity, we 
just concentrate on the the $t$ penguin, $P_t$. We can write
\ber
P_t & = & \bra{M_1,M_2} \sum_i c_i^t O_i\ket{B}\
\eer
Again from the values of the Wilson's coefficients given in Eqn.~\ref{wc}
 we
 can write
\ber
P_t & \approx &  \bra{M_1,M_2} c_3^tO_3+c_4^tO_4 \ket{B}+
\bra{M_1,M_2} c_5^tO_5+c_6^tO_6 \ket{B} \nonumber\\
& - & \frac{1}{2}c_9^t\bra{M_1,M_2} O_9 \ket{B}\nonumber\\
P_t & \approx & c_4^t \bra{M_1,M_2}O_4-(1/N_2)O_3\ket{B}+
c_6^t\bra{M_1,M_2}O_6-(1/N_3)O_5\ket{B} \nonumber\\
& - & \frac{1}{2}c_9^t\bra{M_1,M_2} O_9 \ket{B}\
\label{penguin}
\eer
where from Eqn.~\ref{wc} $N_2=-c_4^t/c_3^t=0.037/0.017=2.2$ and
$N_3=-c_6^t/c_5^t=0.045/0.010=4.5$.

In the diagrammatic approach the $\bar{q'}q'$ 
quarks appearing in the
operator $O_3-O_6$ appears from a ``gluon'' splitting into a 
$\bar{q'}q'$ pair while in the case
the operators $O_7-O_{10}$ it is a ``$ \gamma$'', ``Z'' splitting into a 
$\bar{q'}q'$ pair.

Concentrating on only the term proportional to $c_4^t$, 
we can write in the factorization assumption,
\ber
P_t & = & P_{t1} +P_{t2} \nonumber\\
P_{t1} & = &c_4^t(1-\frac{1}{N_cN_2})\left[
<\rho^0(\omega)|{\bar{u}}\gamma^{\mu}Lb|B^->
<\pi^-|{\bar{d}}\gamma_{\mu}Lu|0>\right.\nonumber\\ 
&+&
\left.<\pi^-|{\bar{d}}\gamma^{\mu}Lb|B^->
<\rho^0(\omega)|{\bar{d}}\gamma_{\mu}Ld|0>\right] \nonumber\\
P_{t2} & = &c_4^t<\pi^-|{\bar{d}}\gamma^{\mu}Lb|B^->
<\rho^0(\omega)|{\bar{u}}\gamma_{\mu}Lu +{\bar{d}}\gamma_{\mu}Ld|0>
[\frac{1}{N_c}-\frac{1}{N_2}] \
\label{penguinrho}
\eer
We see that the second term, $P_{t2}$ has a suppression factor  of 
$(1/N_c-1/N_2) \sim 1/8$. 
This term is called 
 OZI suppressed  and in the 
diagrammatic language this is shown as a ``gluon'' splitting 
into a quark-antiquark pair which then hadronizes to a hadron. Of course 
for a real gluon this process 
is forbidden as the color octet gluon cannot form a color singlet hadron.
 Note that there can be other OZI violating diagrams that have been considered
to explain the large branching ratios in the decay $B \to \eta^{\prime} K$ 
and the semi-inclusive decay
$B \to \eta^{\prime} X_s$. In these diagrams the enhanced branching ratios
are due to the anomaly, gluon couplings to the flavour singlet 
component of the $\eta^{\prime}$ or the intrinsic charm content of the $\eta^{\prime}$
\cite{Atwood:1997bn, Halperin:1998ma}. We will not consider such diagrams 
in our analysis.

The terms represented by $P_{t1}$, in the diagrammatic approach, has a 
``gluon'' splitting into a quark- antiquark pair but now 
the antiquark combines with the
$d$ quark, coming from the $b \to d$ transition, to form a meson 
while the other quark combines with the spectator quark to form 
the second meson in the final state. The two terms 
represented in $P_{t1}$ represent the cases where the
quark-antiquark pair from the ``gluon''  is a $\bar{u} u$ and
a $\bar{d} d$ pair. 

One can do the same exercise with the term proportional to $c_6^t$ in
 Eqn.~\ref{penguin}
 and in this case the OZI violating term is suppressed by
$(1/N_c-1/N_3) \sim 1/9$. Note the term $P_{t2}$ from the
  electroweak penguin term $c_9^t$ 
does not have any suppression. This is expected as a $\bar{q} q$ pair from a
$\gamma$ or $Z$ boson is in a color singlet state and  therefore can form
  a hadron.
 
For the case of $ B^- \to \eta(\eta') K$  
the term $P_{t1}$ and $P_{t2}$ are similar to
the one above.
\ber
P_t & = & P_{t1} +P_{t2} \nonumber\\
P_{t1} & = &c_4^t(1-\frac{1}{N_cN_2})\left[
<\eta(\eta')|{\bar{u}}\gamma^{\mu}Lb|B^->
<K^-|{\bar{s}}\gamma_{\mu}Lu|0>\right. \nonumber\\ 
&+&\left.
<K^-|{\bar{s}}\gamma^{\mu}Lb|B^->
<\eta(\eta')|{\bar{s}}\gamma_{\mu}Ls|0>\right] \nonumber\\
P_{t2} & = &c_4^t<K^-|{\bar{s}}\gamma^{\mu}Lb|B^->
<\eta(\eta')|{\bar{u}}\gamma_{\mu}Lu +
{\bar{d}}\gamma_{\mu}Ld+{\bar{s}}\gamma_{\mu}Ls|0>
[\frac{1}{N_c}-\frac{1}{N_2}] \
\label{penguineta}
\eer

Note we again find the OZI term to be suppressed by a factor $\sim 1/8$. However 
if terms in $P_{t1}$ interfere destructively then the OZI terms may become important. 
Note the suppression in the OZI term can also be diluted if the contributions 
from the $u$,$d$ and $s$ term interfere constructively in $P_{t2}$.
 
In the decay $B^- \to \rho^0(\omega )\pi^-$ the OZI suppressed
 term does not play an important 
role, as these decays are not penguin dominated because 
the CKM factors in the tree and 
penguin terms are of the same order and the penguins are loop suppressed. 
This fact is 
also supported by recent experimental measurement
of the branching ratios 
$ BR[ B^- \to \omega \pi^-] \sim 11.3 \times 10^{-6}$ and 
$BR[ B^- \to \rho^0 \pi^-] \sim 10.4 \times 10^{-6}$ \cite{Jessop:2000bv}. 
Note that in Eqn.~\ref{penguinrho}, from the flavor structure of the $\rho^0$ 
and $\omega$ wavefunction, 
it is obvious 
 that the two terms in in $P_{t1}$ 
interfere destructively for the $\rho^0$ 
but constructively for the $\omega$. Furthermore 
the OZI suppressed term, $P_{t2}$ for 
the $\rho^0$ vanishes, neglecting the electroweak contribution, 
but not for the $\omega$.  
This means that the penguin term for
 $B^- \to \rho \pi^-$ is smaller than in $ B^- \to \omega \pi^-$.
Hence if the penguin terms were dominant then there would be a 
significant difference between the branching ratios
$ BR[ B^- \to \omega \pi^-] $ and 
$BR[ B^- \to \rho \pi^-] $. 
Hence the small measured difference in the  branching ratios for 
$B^- \to \rho^0 \pi^-$ and $B^- \to \omega \pi^-$ implies 
relatively small penguin effects in these decays. 

However decays of the type $B \to \eta K$ are dominated by penguin terms 
because of large CKM factors in the penguin terms compared to the tree term. 
Here, the OZI suppressed terms may have significant 
effects on the predictions of these decays. Note that we expect the OZI
suppressed terms to be more important in $B \to K P$ than in $B \to K V$
decays
 where $P$ is a pseudoscalar and $V$  is a vector state.  
This follows from the fact that
in $J/\psi$ and $\Upsilon$ decays
we know that the OZI-forbidden process requires three gluons for
coupling to a vector meson and two gluons for coupling to a
pseudoscalar. Thus one would expect that the contribution of the OZI
suppressed term should be much smaller in the $B \to K \rho^0(\omega)$ and 
$B \to
K \phi$ decays than in $ B \to K \eta$ and $B \to K \eta^{\prime}$ decays
\cite{Lipkin:1995hn, Lipkin:1997ke}.
One of 
the authors of this work has shown that one can make 
definite predictions about the branching ratios
$ B \to \eta K/B \to \eta^{\prime} K$ and
$ B \to \eta^{\prime} K^*/B \to \eta K^*$ 
\cite{Lipkin:1991us,Lipkin:1992fd} if one 
assumes that the 
OZI terms are forbidden. We will first derive 
these predictions in the language of effective Hamiltonian and 
using the factorization assumption. We then study how the 
predictions change if we use non standard mixing in 
the $\eta-\eta^{\prime}$ sector and if we include the  OZI terms. 

\section{$B \to \rho,\omega, \phi$ transitions.}
In this section we study decays of the type
$B \to V P$ where $V= \rho, \omega, \phi$. As we found in 
section 2 the wavefunction of $V$ has the general form,
\ber
\ket{V} &= & \sum_i a_i \ket{N_i}\
\eer
We can then write, 
\ber
Amp(B \to V M) & = & \sum_i a_i\bra{M,N_i}H_{eff}\ket{B}\
\label{vector}
\eer
 
We now consider the
  ratios of the following decays
\ber
R_{\rho^+} & = & BR({\bar{B}^0} \rightarrow \pi^- \rho^{+ \prime} )/
BR({\bar{B}^0} \rightarrow \pi^- \rho^{+} )
\eer
\ber
R_{\rho^0} & = & BR({\bar{B}^-} \rightarrow \pi^- \rho^{0 \prime} )/
BR({\bar{B}^-} \rightarrow \pi^- \rho^{0} )
\eer
\ber
R_{\omega} & = & BR(B^- \rightarrow \pi^- \omega' )/BR(B^- \rightarrow \pi^- \omega)
\eer
\ber
R_{\phi} & = & BR(B_s \rightarrow \pi^0 \phi' )/BR(B_s \rightarrow \pi^0 \phi)
\eer

As discussed in the previous section we can neglect 
the penguin contribution to these decays. Note that for the decay
$ B_s \rightarrow \pi^0 \phi$ there is no contribution from the
QCD penguin and the
dominant electroweak penguin term has the same structure as the tree amplitude
 and hence the ratio $R_{\phi}$ remains essentially the
 same even in the presence of penguin terms.
We then obtain
\ber
R_{\rho^+}  & = & \left|\frac{\bra{\rho^{+ \prime}}\bar{u}\gamma^{\mu}(1-\gamma_5)b\ket{\bar{B}^0}
\bra{\pi^-}\bar{d}\gamma_{\mu}(1-\gamma_5)u\ket{0}}
{\bra{\rho^{+ }}\bar{u}\gamma^{\mu}(1-\gamma_5)b\ket{\bar{B}^0}
\bra{\pi^-}\bar{d}\gamma_{\mu}(1-\gamma_5)u\ket{0}}\right |^2 \nonumber\\
& = & \left |\frac{A_0^{\prime}}{A_0}\right |^2\frac{P_{\rho^{+ \prime}}^3}
{P_{\rho}^3}\
\eer
where $P$ is the magnitude of the three momentum of the final states and the form factor $A^0$ is defined through
\ber
\bra{V_f}A_{\mu}\ket{P_i} & = &
(M_i +M_f)A_1\left[\epsilon_{\mu}^* -\frac{\epsilon^*.q}{q^2}q_{\mu}\right]
-A_2 \frac{\epsilon^*.q}{M_i+M_f}\left[
(P_i+P_f)_{\mu} -\frac{M_i^2-M_f^2}{q^2}q_{\mu}\right] \nonumber\\
& + & 2M_f A_0 \frac{\epsilon^*.q}{q^2}q_{\mu}\
\eer
\ber
R_{\rho^0}  & \approx & \left|\frac{\bra{\rho^{0 \prime}}\bar{u}\gamma^{\mu}(1-\gamma_5)b\ket{\bar{B}^0}
\bra{\pi^-}\bar{d}\gamma_{\mu}(1-\gamma_5)u\ket{0}}
{\bra{\rho^{0 }}\bar{u}\gamma^{\mu}(1-\gamma_5)b\ket{\bar{B}^0}
\bra{\pi^-}\bar{d}\gamma_{\mu}(1-\gamma_5)u\ket{0}}\right |^2 \nonumber\\
& = & \left|\frac{A_0^{\prime}}{A_0}\right |^2\frac{P_{\rho^{0 \prime}}^3}{P_{\rho^0}^3}\
\eer

In the above we have ignored the term in the amplitude for 
$ B^- \to \rho(\rho^{\prime}) \pi^-$ which is given by $T_1$ in Eqn.~\ref{tree}
 and hence is suppressed by $ \sim 1/15$ relative 
to the dominant term $ T_2$.

Within the same approximation we can then write
\ber
R_{\omega}  & \approx & \left|\frac{\bra{\omega^{ \prime}}\bar{u}\gamma^{\mu}(1-\gamma_5)b\ket{\bar{B}^0}
\bra{\pi^-}\bar{d}\gamma_{\mu}(1-\gamma_5)u\ket{0}}
{\bra{\omega}\bar{u}\gamma^{\mu}(1-\gamma_5)b\ket{\bar{B}^0}
\bra{\pi^-}\bar{d}\gamma_{\mu}(1-\gamma_5)u\ket{0}}\right|^2 \nonumber\\
& = & \left|\frac{A_0^{\prime}}{A_0}\right|^2\frac{P_{\omega^{ \prime}}^3}{P_{\omega}^3}\
\eer
Finally,
\ber
R_{\phi}  & = & \left|\frac{\bra{\phi{ \prime}}\bar{s}\gamma^{\mu}(1-\gamma_5)b\ket{\bar{B}_s}
\bra{\pi^0}\bar{u}\gamma_{\mu}(1-\gamma_5)u\ket{0}}
{\bra{\phi}\bar{s}\gamma^{\mu}(1-\gamma_5)b\ket{\bar{B}^0}
\bra{\pi^0}\bar{u}\gamma_{\mu}(1-\gamma_5)u\ket{0}}\right|^2 \nonumber\\
& = & \left|\frac{A_0^{\prime}}{A_0}\right|^2\frac{P_{\phi{ \prime}}^3}
{P_{\phi}^3}\
\eer

To calculate the ratios we need the form factor $A_0$. We use the  
model for form factors in Ref\cite{Aleksan:1995bh} 
that incorporates 
some relativistic features
and is relatively simple to use. The calculation of 
the form factors in this model
is not always in agreement with experimental results, however, we expect 
the model
to make reliable predictions for the ratio of form factors. We believe 
this feature
to be true for most other models for form factors. Hence we 
shall mostly calculate
quantities that can be expressed as ratios of form factors. The model 
in \cite{Aleksan:1995bh}
assumes, like many  other models, the weak binding limit 
which sets the meson mass to be the sum of the 
masses of the constituent quarks making up the meson. Hence the effect of
binding are included in the quark masses. This is a good approximation for the
lowest resonances of the $\rho, \omega$ and the $\phi$ systems.
In this model the form factor $A_0$ is given by
\bers
A_0 & = & A[J_1-\frac{M_{-}}{M_{+}}J_2]\nonumber\\
A & = & \frac{\sqrt{4M_iM_f} M_{+}}{M_{+}^2-q^2}\nonumber\\
J_1 & = & \int d^3p \phi_f^{*}(\vec p + \vec a) \phi_i(\vec p) \nonumber\\
J_2 & = & m_s\int d^3p \phi_f^{*}(\vec p + \vec a) \phi_i(\vec p)
[\frac{\vec p. \vec a}{\mu a^2} + \frac{1}{m_f}]\
\eers
where
\ber
M_{\pm} & = & M_f \pm M_i \nonumber\\
\vec a & = &2 m_s \vec{\beta} =2 \frac{m_s \tilde {q}}{M_{+}} \nonumber\\
\tilde{q}^2 & =& M_{+}^2 \frac{M_{-}^2-q^2}{M_{+}^2-q^2} \nonumber\\
\mu & = & \frac{m_im_f}{m_i+m_f} \
\label{model1}
\eer
and $\phi_f$ and $\phi_i$ represent the momentum space wave functions while 
$\vec{\beta}$
is the velocity of the mesons in the equal velocity frame and $m_{i,f}$
are the non spectator quark masses of the initial and final meson.

We use the momentum wavefunction $\phi_f$ obtained from spectroscopy in 
section. (2)
while for $\phi_i$ we use the wave function
\ber
\phi_i& = & \phi_B=N_B e^{-p^2/p_F^2} \
\label{phiB}
\eer
where $p_F$ is the Fermi momentum of the B meson. In our calculations we 
will take
$p_F=300$MeV. Note that in the analysis presented in the introduction we 
have
neglected the  Fermi momentum of the b quark, but since
$p_F/m_b$ is small  the general conclusions 
reached in the introduction still continue to be valid.  

For transitions to higher resonant states,
we use the same quark masses as those used in the
transition of the $B$ meson to the lowest resonant state. This is 
reasonable, as
 the spectator quark still comes from the
 $B$ meson and therefore has the same value for its mass 
irrespective of whether the final state is in the lowest or 
the first excited state. The values for the masses of the 
non spectator masses are taken to be  essentially the same 
as those used in section. (2) for spectroscopy. However for 
the calculation of the velocity $\vec{\beta}$ and hence $\vec{a}$ defined 
in Eqn.~\ref{model1} we use the physical mass of the higher resonant state.

\begin{table}[thb]
\caption{Ratios of branching ratios for different confining potentials}
\begin{center}
\begin{tabular}{|c|c|c|c|}
\hline
Ratio & Linear & Quadratic & Quartic  \\
\hline
$R_{\rho^+}$ &
$2.3$ & $2.0$ & $1.9$ \\
\hline 
$R_{\rho^0}$ &
$2.3$ & $2.0$ & $1.9$ \\
\hline
$R_{\omega}$ &
$3.5$ & $2.5$ & $1.7$ \\
\hline
$R_{\phi}$ &
$6.7$ & $6.2$ & $5.2$ \\
\hline 
\end{tabular}
\end{center}
\label{T1}
\end{table}
In  Table.~\ref{T1}
 we give our predictions for the various ratios defined above.
 We find that the transitions
to higher excited states can be comparable or enhanced relative to the 
transitions to the ground state. From Table.~\ref{T1} 
we see that
the ratios of branching ratios are slightly sensitive 
to the confining potential 
and 
the ratios of branching ratios increase as we go from the quartic to 
the linear potential. This is because the wavefunction 
for the linear potential has a longer tail and hence more 
high momentum components than the wavefunction 
for the quadratic  and the quartic potentials. The wavefunction 
for the quadratic potential, has in turn,
 a longer tail and hence more 
high momentum components than the wavefunction for the quartic potential. 
As mentioned in section 1
the form factor in $B \to M$ transition, where $M$ is a light meson, 
is sensitive to the high momentum tail of the meson wavefunction.
Hence we would expect the hierarchy
$(A_{1,0})_{linear} >(A_{1,0})_{quadratic} >(A_{1,0})_{quartic} $ 
where $A_{1,0}$ are the form factors for the transition of $B$ to the 
first radially excited and the ground state of the meson $M$.
We see from from Table.~\ref{T1} that this
hierarchy 
is maintained for the ratios of form factors and so we have
$(A_{1}/A_{0})_{linear} >(A_{1}/A_{0})_{quadratic} > 
(A_{1}/A_{0})_{quartic}  $.
Note that the ratio of form factors also depend on the choice of the 
Fermi momentum of the $B$ meson, as 
a smaller(larger) Fermi momentum would
make the form factors more(less) sensitive to 
the tail of the wavefunction of $M$, as well as mixing effects in 
the wavefunction of the meson $M$.

One can check that the predictions for the ratios of branching ratios 
 are not very different from the case where we 
neglect radial mixings. Hence
 one concludes  
  that the effects of radial mixing are in most cases small 
and negligible for practical applications.  
 We observe in Table.~\ref{T1}  
  that there can be a large enhancement for $R_{\phi}$.
One can get a rough estimate of the branching ratio for $B_s \to \phi \pi^0$ 
from
\bers
\frac{BR[B_s \to \phi \pi^0]}{BR[\bar{B}^0 \to \rho^+ \pi^-]} & \approx &
\frac{1}{2} \left|\frac{V_{ub}V_{us}^{*}(c1+c2/N_c) -V_{tb}V_{ts}^* c_9/2}
{V_{ub}V_{ud}^{*}(c2+c1/N_c)}\right|
\approx 0.04 \
\eers
where we have neglected form factor  and phase space differences between
$B_s \to \phi \pi^0$ and 
$\bar{B}^0 \to \rho^+ \pi^-$. Using the measured
$BR[ B \to \rho^{+} \pi^-] \sim 28 \times 10^{-6}$ \cite{Jessop:2000bv}  we get
$BR[B_s \to \phi \pi^0]  \sim 10^{-6}$. Hence the large enhancement
 for $R_{\phi}$ indicates that
  $BR[B_s \to \phi^{\prime} \pi^0]$ can be of $O(10^{-5})$.

Note that in the $\rho(\omega)$ system there are two resonances, 
$\rho(1450)[\omega(1420)]$ 
and $\rho(1700)[\omega[1650]$, which can be identified 
a S-wave radial excitation(2S) and a 
D wave orbital excitation in the quark model. However 
recent studies of the decays of 
these resonances show that it is possible that these states are mixtures of
$q \bar{q}$ and hybrid 
states Ref\cite{rpp2000}. Hence the state $\rho(1450)[\omega(1420)]$  
is interpreted as
a 2S state with a small mixture of a hybrid state. We do not take 
into account such possible mixing with a hybrid state in our 
calculation and the meson masses
for these excited states used in our calculation are the ones 
we predict in section 2.  For the $\phi$ system
there is only state at $\phi(1680)$ which we interpret as a 2S state in the absence of mixing effects.

\section {\bf $B^{\pm}\to K^{\pm}(K^{* \pm}) \eta^{\prime}(\eta)$}
We construct two ratios
\ber
R_{K} & = & 
BR(B^- \rightarrow K^- \eta )/BR(B \rightarrow K^- \eta^{\prime})
\label{bk}
\eer
 and
\ber
R_{K^*} & = & 
BR(B^- \rightarrow K^{-*} \eta^{\prime} )/BR(B \rightarrow K^{-*} \eta)
\label{bkstar}
\eer
Let us assume the $\eta-\eta^{\prime}$ mixing used in Ref\cite{Lipkin:1991us}
\ber
\ket{\eta} & = & \frac{1}{\sqrt{2}}\left[N_0 -S_0 \right] \nonumber\\
\ket{\eta^{\prime}} & = & \frac{1}{\sqrt{2}}\left[N_0 +S_0 \right] \
\label{Isgur}
\eer
where, as before,
$|N>=| u \bar{u} + d \bar{d}>/\sqrt{2}$ and $S=|s \bar{s}>$.
Now, from Eqn.~\ref{penguineta}  if we only include the term $P_{t1}$ then we find
\ber
R_{K} & \approx & \left|\frac{f_KF^{+}_{\eta}+
f_{\eta}^sF^{+}_{K}}
{f_kF^{+}_{\eta^{\prime}}+
f_{\eta^{\prime}}^sF^{+}_{K}}\right|^2\
\label{rk}
\eer
where we have dropped the masses of the pseudoscalars in the final states and
 the form factor $F^{+}$ is defined through 
\ber
\bra{P(p_f)}\bar{q} \gamma_\mu (1-\gamma_5) b \ket{ B(p_i) }
&=& \left[ (p_i+ p_{f})_\mu - \frac{m_B^2-m_{P}^2}{q^2} 
q_\mu \right]
F_1(q^2)\nonumber\\
& + & \frac{m_B^2-m_{P}^2}{q^2} q_\mu F_0(q^2) \nonumber\\
&=& F^{+}_{P}(p_i+ p_{f})_\mu +F^{-}_{P}q_{\mu}. \
\label{ffactor}
\eer
In the above equation $f_K$ is the kaon decay constant and the 
decay constants $f_{\eta}^q$ and $f_{\eta^{\prime}}^q$ are defined by,
\ber
if_{\eta(\eta^{\prime})}^q p^{\mu}_{\eta(\eta^{\prime})} & = & \bra{\eta(\eta^{\prime})}
\bar{q}\gamma^{\mu}(1-\gamma_5) q \ket{0}.\
\label{decayconstant}
\eer 

For the mixing in Eqn.~\ref{Isgur}, and assuming  $SU(3)$ flavor symmetry, we can write
\ber
f_{\eta}^{u,d} \approx \frac{f_K}{2}\nonumber\\
f_{\eta}^s \approx \frac{-f_K}{\sqrt{2}}\nonumber\\
f_{\eta^{\prime}}^{u,d} \approx \frac{f_K}{2}\nonumber\\
f_{\eta^{\prime}}^s \approx \frac{f_K}{\sqrt{2}}\nonumber\\
F_{\eta}^{+} \approx \frac{F_K^{+}}{{2}}\nonumber\\
F_{\eta^{\prime}}^{+} \approx \frac{F_K^{+}}{{2}}\
\label{fsymmetry}
\eer
One can then write
\ber
R_K & \approx & \left|\frac{\frac{1}{2}-\frac{1}{\sqrt{2}}}
{\frac{1}{2}+\frac{1}{\sqrt{2}}}\right|^2 \sim 0.03 \
\label{rkn}
\eer
It was shown in Ref\cite{Lipkin:1991us,Lipkin:1992fd} that there is a parity selection 
rule in the decays $ B \to \eta(\eta^{\prime})K^{(*)}$ which fixes the
relative phase between the  penguin amplitudes representing the
strange and nonstrange contributions  to $\eta$ and $\eta^{\prime}$ final states.
 In particular the parity selection rule predicts that the 
phases between the strange and nonstrange penguin amplitudes 
in $B \to K \eta(\eta^{\prime})$ and
$B \to K^* \eta(\eta^{\prime})$ are reversed. Hence
neglecting form factor differences for $B \to P$ and $B \to V$ transitions, one obtains 
\ber
R_{K^*} & \approx & \left|\frac{\frac{1}{2}-\frac{1}{\sqrt{2}}}
{\frac{1}{2}+\frac{1}{\sqrt{2}}}\right|^2 \sim 0.03 \
\label{rkstarn}
\eer

We now calculate the ratios above with 
a nonstandard $\eta-\eta^{\prime}$
mixing.
We will use the mixing for a quartic potential given in Eqn.~\ref{quartic}. 
This is because the ground states of this 
mixing has the largest overlap with the standard mixing in Eqn.~\ref{Isgur}.
Later on in this paper we will compare form factors calculated with the
mixing in Eqn.~\ref{quartic} with those calculated with the mixing in Eqn.~\ref{Isgur} 
to get an idea of the effects of radial mixing  in the 
$\eta-\eta^{\prime}$ wavefunctions. A key ingredient in the parity selection rule
that predicts the 
relative phases of the
strange and nonstrange contributions  to $\eta$ and $\eta^{\prime}$ final states is
approximate flavor symmetry. In the flavor symmetry limit the radial
wave functions of $\pi$, K, $\eta$
and $\eta^{\prime}$ ( up to mixing factors in $\eta(\eta^{\prime}$)  
are all the same. One can then argue that since
all  are  ground
state wave functions with no nodes and a constant phase over the entire
radial   domain,
flavor symmetry breaking can change the radial shape and size of the
wave function but will not reverse the phases. This is probably  
a reasonable assumption
as long as we only have
ground state wave functions. But when radially excited wave functions
which have nodes are also considered, there is a phase reversal in the
wave function.
 Note that with factorization,
the B decay into the nonstrange part of the $\eta$ and 
$\eta^{\prime}$ involves a
point like form factor for the kaon and and a
hadronic overlap integral for the $\eta (\eta^{\prime})$. 
But
the B decay into the strange part of the $\eta$ and $\eta^{\prime}$ involves a
point like form factor for the $\eta (\eta^{\prime})$  
and a hadronic overlap integral
for the kaon.
The penguin amplitude that involves a hadronic overlap
integral for the $\eta (\eta^{\prime})$ can cause
a phase ambiguity between 
the two penguin amplitudes because the
wave function for the radial excitation changes phase and 
we are not able to 
 exactly calculate flavor breaking effects.

A simple model to simulate flavor symmetry breaking
would be to define, for the
$\eta(\eta^{\prime})$, an effective wavefunction 
\ber
\Psi_{eff} & = & a_0 \Psi_{N_0} + a_1 r_1 \Psi_{N_1} +a_2 r_2 \Psi_{N_2}\
\label{ewavefunction}
\eer
where the $a_i$'s are from Eqn.~\ref{quartic} and $r_1$ and $r_2$, representing flavor 
breaking effects, 
can have both signs. The effective wavefunction
$\Psi_{eff}$ will then be used for form factor calculations. 
We will consider two choices of $r_1$ and $r_2$. In the first case we set
$r_1=r_2=1$ and in the second case we choose $r_1$ and $r_2$ to be such that
the contributions from the various radial excitations add constructively in
the
form factor calculations involving the $\eta$ and the $\eta^{\prime}$.
    
The form factors  in Eqn. ~\ref{ffactor}  calculated using the quark model of 
Ref\cite{Aleksan:1995bh} are given by
\ber
F_{+} & = & A[J_1+\frac{M_{-}}{M_{+}}J_2]\nonumber\\
F_{-} &= & A[J_2+\frac{M_{-}}{M_{+}}J_1]\nonumber\\
A & = & \frac{\sqrt{4M_iM_f} M_{+}}{M_{+}^2-q^2}\nonumber\\
J_1 & = & \int d^3p \phi_f^{*}(\vec p + \vec a) \phi_i(\vec p) \nonumber\\
J_2 & = & m_s\int d^3p \phi_f^{*}(\vec p + \vec a) \phi_i(\vec p)
[\frac{\vec p. \vec a}{\mu a^2} + \frac{1}{m_f}]\
\eer
where
\ber
M_{\pm} & = & M_i \pm M_f \nonumber\\
\vec a & = &2 m_s \vec{\beta} =2 \frac{m_s \tilde {q}}{M_{+}} \nonumber\\
\tilde{q}^2 & =& M_{+}^2 \frac{M_{-}^2-q^2}{M_{+}^2-q^2} \nonumber\\
\mu & = & \frac{m_im_f}{m_i+m_f} \
\eer
and $\phi_f$ and $\phi_i$ represent the momentum space wave functions, 
$\vec{\beta}$
is the velocity of the mesons in the equal velocity frame and $m_{i,f}$ are the
non spectator quark masses of the initial and final meson.
For the calculation of the form factors in the $\eta(\eta^{\prime})$ system, 
clearly, the weak binding assumption, which would imply 
$M_{\eta} \sim M_{\eta^{\prime}} \sim M_{\rho}$, does not hold. One
does not know how to incorporate corrections to the weak binding limit.
One could choose, for instance, different quark masses, and  
there are various reasonable choices of quark masses that can be made leading 
to very different predictions for the form factors. We do not wish 
to explore all these possibilities here, instead we will use 
the same procedure
and the same
values for the  quark masses   
used in the calculation of the form factors
in the vector  system. We then find
\ber 
\frac{F_{+nonstandard}^{\eta^{\prime}}}{F_{+standard}^{\eta^{\prime}}}
&\approx & 1.5,1.7\nonumber\\
\frac{F_{+nonstandard}^{\eta}}{F_{+standard}^{\eta}}
&\approx & 0.5, 2.1\
\label{formfactor1}
\eer
where $F_{+standard}$ and $F_{+nonstandard}$ are the form factors calculated in 
the standard mixing in Eqn.~\ref{Isgur}, and for the 
nonstandard mixing in Eqn.~\ref{quartic}. The contribution 
from the form factor $F_{-}$ is negligible.
The two numbers in the equation above correspond to 
the two choices  for $r_{1,2}$ 
mentioned above. We note that the form factor 
with nonstandard mixing does not change 
much for the $\eta^{\prime}$ but changes significantly for the $\eta$
for the two choices of $r_{1,2}$.  

Using the mixing in Eqn.~\ref{quartic} one obtains for the
decay constants
\ber
\frac{f^{u,d}_{{\eta}_{nonstandard}}}{f^{u,d}_{{\eta}_{standard}}} & \approx & 1\nonumber\\
\frac{f^{u,d}_{{\eta^{\prime}}_{nonstandard}}}
{f^{u,d}_{{\eta^{\prime}}_{standard}}} & \approx & 1.1\nonumber\\
\frac{f^{s}_{{\eta}_{nonstandard}}}{f^{s}_{{\eta}_{standard}}} & \approx 
& 0.8\nonumber\\
\frac{f^{s}_{{\eta^{\prime}}_{nonstandard}}}{f^{s}_{{\eta^{\prime}}_
{standard}}} & \approx & 1.2\
\label{dconstant1}
\eer

We will choose the second entry in Eqn.~\ref{formfactor1} and for simplicity
use
\ber
\frac{F_{+nonstandard}^{\eta^{\prime}}}{F_{+standard}^{\eta^{\prime}}}
&\approx & 2.0 \approx
\frac{F_{+nonstandard}^{\eta}}{F_{+standard}^{\eta}}\nonumber\\
{f^{u,d,s}_{{\eta}_{nonstandard}}}& \approx & {f^{u,d,s}_{{\eta}_{standard}}} 
\nonumber\\
{f^{u,d,s}_{{\eta^{\prime}}_{nonstandard}}} & \approx &
{f^{u,d,s}_{{\eta^{\prime}}_{standard}}} \
\label{new}
\eer
The relations in Eqn.~\ref{new} are within the uncertainties in the 
calculation of the form factors
in Eqn.~\ref{formfactor1} and Eqn.~\ref{dconstant1}

It is now easy to check that the predictions
for $R_{K}$ and $R_{K^*}$ in Eqn.~\ref{rkn} and
 Eqn.~\ref{rkstarn} 
remain essentially
 unchanged. However, 
as was shown by one of the authors of this paper, one can use flavour  
topology characteristics of 
charmless $B$ decays to derive additional sum rules  connecting 
$B$ decays to $K \eta({\eta^{\prime}})$ and $ K \pi$ final states.
One of the interesting sum rule derived in 
\cite{Lipkin:1997ad, Lipkin:1997ke, Lipkin:2000sf} is, neglecting 
phase space corrections,
\ber
R & = & \frac{\Gamma[B^{\pm} \to K^{\pm} \eta^{\prime}]
+
\Gamma[B^{\pm} \to K^{\pm} \eta]}
{\Gamma[B^{\pm} \to K^{\pm} \pi^0]}  \le  3 \
\label{sumrule}
\eer
Note that this sum rule is true for any standard $\eta-\eta^{\prime}$
mixing.
Recent experimental measurements \cite{Chen:2000hv}, however, show the above 
sum rule to be invalid.
In the factorization assumption, one can write, neglecting the
 tree contribution,
\ber
R & \approx & \frac{\left|f_KF^{+}_{\eta}+
f_{\eta}^sF^{+}_{K}\right|^2+
\left|f_KF^{+}_{\eta^{\prime}}+
f_{\eta^{\prime}}^sF^{+}_{K}\right|^2}{\left|F^{+}_{\pi^0}f_K\right|^2}.
\eer
With the mixing in Eqn.~\ref{Isgur} we get $R \approx $ 3. Note that  
 one can check that
for any standard $\eta-\eta^{\prime}$
mixing one always gets  $R \approx $ 3. 
With the
  nonstandard $\eta-\eta^{\prime}$
mixing in Eqn.~\ref{quartic},
 we get $R \approx $ 6 which is now consistent with experiment.
 One would also get similar predictions with a $K^*$ in the final state.
In particular we have
\ber
\frac{\Gamma[B^{\pm} \to K^{*\pm} \eta]}
{\Gamma[B^{\pm} \to K^{*\pm} \pi^0]}  & = &  |\sqrt{2} +1|^2 \approx 6
\nonumber\\
\frac{\Gamma[B^{\pm} \to K^{*\pm} \eta^{\prime}]}
{\Gamma[B^{\pm} \to K^{*\pm} \pi^0]}  & = &  |\sqrt{2} -1|^2 \approx 
\frac{1}{6}
\
\eer

As we have argued before the OZI suppressed terms may play an 
important role in the decays $ B \to \eta(\eta^{\prime}) K^{(*)}$ decays. In 
particular the OZI suppressed terms are important for decays with 
a $\eta^{\prime}$ in the final state because they add constructively while
for the $\eta$ in the final state the OZI suppressed terms tend to cancel 
among 
themselves.
One can calculate the contribution to the amplitude
 of the OZI suppressed term with 
the mixing in Eqn.~\ref{Isgur} as, 
\ber
x_{\eta^{\prime}} & \approx & \frac{1}{8}(1+\frac{1}{\sqrt{2}}) \sim 0.21 
\nonumber\\
x_{\eta} & \approx & \frac{1}{8}(1-\frac{1}{\sqrt{2}}) \sim 0.037\
\eer
where we have dropped factors common to both $x_{\eta}$ and 
$x_{\eta^{\prime}}$.
We see that  the OZI suppressed contribution for the $\eta^{\prime}$ in the final state
is indeed more important than for the $\eta$ in the final state.

We now present the full amplitude
for the decays $ B \to \eta(\eta^{\prime})K(K^{*})$ 
including all the terms in the effective Hamiltonian
in the factorization assumption and including the OZI suppressed terms.
To make definite predictions we choose the  form factors, 
$F_{+standard}^{\eta(\eta^{\prime})}$, as well 
as the form factors for $B \to K(K^*)$ transitions to be given by 
Ref\cite{BSW}. We will use the decay constants
$f_{\eta}^{u,d}=f_{\eta^{\prime}}^{u,d} \approx 0.8 f_{\pi}$ and
$f_{\eta^{\prime}}^{s}=-f_{\eta}^s \approx 1.3 f_{\pi}$ .

Finally, one can write the amplitude 
for $ B^- \to K^- \eta'$   as \cite{Datta:1998nr}
\bers
M & = & \frac{G_F}{\sqrt 2}
\left[V_u \left(a_1 r_1 Q_K + a_2 Q_{\eta'} \right)
-\sum_{i=u,c,t}V_i \left\{(T_1^i r_1 +T_2^i r_2)Q_K +T_3^i
Q_{\eta'}\right\}\right]\
\eers
where
\ber
T_1^i &= & 2a_3^i- 2a_5^i- \frac{1}{2}a_7^i + \frac{1}{2}a_9^i\nonumber\\
T_2^i & = & a_3^i +a_4^i- a_5^i
+(2a_6^i -a_8^i)
\frac{m_{\eta^\prime}^2}{2m_s(m_b-m_s)} + \frac{1}{2}a_7^i 
- \frac{1}{2}a_9^i - \frac{1}{2}a_{10}^i\nonumber\\
T_3^i & = & a_4^i + 2(a_6^i +a_8^i)
\frac{m_K^2}{m_u+m_s} \frac{1}{m_b-m_u} 
+a_{10}^i \
\label{fack}
\eer
with 
$a_1 = c_1 + \frac{c_2}{N_c} $,
$ a_2 = c_2 + \frac{c_1}{N_c} $,
$a^i_j = c^i_j + \frac{c^i_{j+1}}{N_c} $, 
$a^i_{j+1} = c^i_{j+1} + \frac{c^i_j}{N_c}$, 
$r_1 = \frac{f_{\eta^\prime}^u}{f_\pi}$
$r_2 = \frac{f_{\eta^\prime}^s}{f_\pi} $,
$Q_K = i F_0^K (m_{\eta^\prime}^2) (m_B^2 - m_K^2) f_\pi$,
$Q_{\eta^\prime} = i F_0^{\eta^\prime} (m_K^2)
(m_B^2 - m_{\eta^\prime}^2) f_K $, 
$V_i = V_u, V_c, V_t$ and $N_c$ is effective number of colors.

In the above equations we have used the quark equations of motion to 
simplify certain matrix
elements. The masses used in these equations of motion are the current quark masses given in section 3.
The expression for the amplitude can also be used for
$B\rightarrow \eta K$ by making the necessary changes. It is also
straight forward to write down the amplitudes for
$B\rightarrow K^{*} \eta^\prime$ and $B\rightarrow K^{*} \eta$ decays.
\bers
M & = & \frac{G_F}{\sqrt 2}
\left[V_u \left(a_1 f_{\eta'}^u A + a_2 m_{K^*}g_{K^*} B \right)
-\sum_{i=u,c,t}V_i \left\{(S_1^i f_{\eta'}^u +
S_2^i f_{\eta'}^s )A +S_3^i
m_{K^*}g_{K^*} B\right\}\right]\
\eers
where
\ber
S_1^i &= & 2a_3^i- 2a_5^i- \frac{1}{2}a_7^i + \frac{1}{2}a_9^i\nonumber\\
S_2^i & = & a_3^i +a_4^i- a_5^i
-(2a_6^i -a_8^i)
\frac{m_{\eta^\prime}^2}
{2 m_s (m_b + m_s)} + \frac{1}{2}a_7^i 
- \frac{1}{2}a_9^i - \frac{1}{2}a_{10}^i\nonumber\\
S_3^i & = & a_4^i +a_{10}^i \
\label{fackstar}
\eer

with
$A = 2m_{K^*}A_0 \varepsilon^*\cdot p_B $,
$B = 2 \varepsilon^{ *} \cdot p_B
F_1^{\eta^\prime} (m_{K^*}^2)$ 
and we will use $g_{K^{*}} =221 $ MeV where $g_{K^{*}}$ is 
the vector meson decay constant.
A similar expression for $B\rightarrow \eta K^{*}$ can also be obtained.
To identify the various terms in Eqn.~\ref{fack} and
Eqn.~\ref{fackstar}
, let us for simplicity drop the electroweak penguin terms represented by $a_7^i-a_{10}^i$. Note that the
term proportional to $a_6^i$ is formally of $O(1/m_b)$ and so strictly in the 
$m_b \to \infty$ limit this term vanishes. However for 
realistic quark masses this term is not negligible and is chirally 
enhanced because of the strange quark mass, $m_s$
which is given in section 3. The OZI suppressed
 terms are represented by the terms $a_3^i$ and $a_5^i$. One can check that if we drop the OZI suppressed
 terms as well as the chirally enhanced terms then, 
neglecting form factor and phase space differences, 
we would recover the predictions of Eqn.~\ref{rkn}
and Eqn.~\ref{rkstarn} from 
Eqn.~\ref{fack} and
 Eqn.~\ref{fackstar}.
If we include the OZI suppressed terms in the calculation  
without the chirally enhanced
contribution we obtain
$BR[ B \to \eta^{\prime}K^{*}] =1.1 \times 10^{-6}$ while if we 
ignore the OZI suppressed terms then we obtain
$BR[ B \to \eta^{\prime}K^{*}] =0.34 \times 10^{-6}$. 
Hence we see that the presence of the OZI suppressed
 term can alter significantly the 
$BR[ B \to \eta^{\prime}K^{*}]$. There is a much smaller effect of the 
OZI suppressed terms
in $BR[ B \to \eta^{\prime}K]$. This, as already mentioned, is due to 
the fact that the  OZI allowed terms in
$BR[ B \to \eta^{\prime}K]$ 
tend to add constructively while they add
destructively in 
$BR[ B \to \eta^{\prime}K^{*}]$  
and so the OZI suppressed effects are
felt more strongly in $ B \to \eta^{\prime}K^{*}$.  
\begin{table}[thb]
\caption{Branching ratios(BR) for $B \to \eta(\eta')K(K^*)$ decays }
\begin{center}
\begin{tabular}{|c|c|c|}
\hline
Process & Experimental BR\cite{Richichi:2000kj} & Theory BR \\
\hline
$B^-\rightarrow K^- \eta'$ &
$(80 ^{+10}_{-9} \pm 7)\times 10^{-6}$ &$ 93\times 10^{-6}$ \\
\hline 
$B^-\rightarrow K^- \eta$ &
 $ < 6.9 \times 10^{-6}$ 
&$ 1.04 \times 10^{-6}$ \\
\hline
$B^-\rightarrow K^{-*} \eta'$ &
 $ < 35\times 10^{-6}$ 
&$ 3.6 \times 10^{-6}$ \\
\hline 
$B^-\rightarrow K^{-*} \eta $ &
$(26.4 ^{+9.6}_{-8.2} \pm 3.3)\times 10^{-6}$ &$ 10 \times 10^{-6}$ 
\\ \hline 
\end{tabular}
\end{center}
\label{T2}
\end{table}
Finally we present
in the Table.~\ref{T2} the results of our calculation including 
all the terms in
Eqn.~\ref{fack} and
 Eqn.~\ref{fackstar}.
We see from Table.~\ref{T2}  that our calculations are in reasonable 
agreement with experiment. In particular we note that the chirally enhanced contributions tend to further increase the branching ratio
$BR[ B \to \eta^{\prime}K^{*}]$. 
 From the table above we can calculate
\ber
R_K & = & 0.01 \nonumber\\
R_{K^*} & = & 0.36 \
\eer
If experiments find that the ratio $R_{K^*}$ is indeed much smaller 
than predicted here then this would indicate the presence of
 large non factorizable corrections
that effectively cancel the OZI contributions as well as the chirally enhanced corrections.
\section{Summary}
We have considered the weak decays of a B meson to final states 
that are mixtures 
of S-wave radially excited components. We calculated nonleptonic 
decays of the type $B \to \rho' \pi/B \to \rho \pi$, 
$B \to \omega' \pi/B \to \omega \pi$ and
$B \to \phi' \pi/B \to \phi \pi$ 
where $\rho'$, $\omega'$ and $\phi'$ are higher $\rho$, $\omega$ 
and $\phi$ resonances. We found that the transitions to the excited
states can be comparable or enhanced relative to transitions 
to the ground state. It would, therefore, be 
extremely interesting to test these 
predictions.
We also studied the effect of radial mixing in the vector and 
the pseudoscalar systems generated from hyperfine interaction 
and the annihilation
term. We found the effects of radial mixing to be small and
 generally negligible for all practical purposes in the vector system.
However, in  the $\eta-\eta^{\prime}$ system the effects 
of radial mixing are appreciable and
seriously affect decay branching ratios. In particular we found that
the experimental violation of the sum rule for $B\to K \eta^{\prime}$ in 
Eqn.~\ref{sumrule} can be explained by radial 
mixing without need for the OZI suppressed transitions. We also pointed 
out that
the place to look for an OZI suppressed contribution is in
$B \to K^* \eta^{\prime}$ decays where the   
 the OZI suppressed transitions become important as 
OZI allowed term is  small.
\centerline{ {\bf  Acknowledgment}}
This work was  supported  by the
US-Israel Bi-National Science Foundation
 and by the Natural  Sciences and Engineering Research  Council
of Canada.


\begin{thebibliography}{99}
\bibitem{Cohen:1979ge}
I.~Cohen and H.~J.~Lipkin,
Nucl.\ Phys.\ {\bf B151}, 16 (1979).

\bibitem{Frank:1984bj}
M.~Frank and P.~J.~O'Donnell,
Phys.\ Rev.\ D {\bf 29}, 921 (1984).


\bibitem{Okubo:1963fa}
S.~Okubo,
Phys.\ Lett.\ {\bf 5}, 165 (1963).
J.~Iizuka,
Prog.\ Theor.\ Phys.\ Suppl.\ {\bf 37-38}, 21 (1966).

\bibitem{Zweig}
G. Zweig, CERN Report No 8419/TH 414 (1961).

\bibitem{Quigg:1979vr}
C.~Quigg and J.~L.~Rosner,
Phys.\ Rept.\ {\bf 56}, 167 (1979).




\bibitem{Bramon:1974ky}
A.~Bramon,
Phys.\ Lett.\ {\bf B51}, 87 (1974).

\bibitem{Lipkin:1997ad}
H.~J.~Lipkin,
Phys.\ Lett.\ {\bf B415}, 186 (1997)
[hep-ph/9710342].

\bibitem{Reina} 
A.~J.~Buras, M.~Jamin, M.~E.~Lautenbacher and P.~H.~Weisz,
Nucl.\ Phys.\ {\bf B400}, 37 (1993)
[hep-ph/9211304].
A.~J.~Buras, M.~Jamin and M.~E.~Lautenbacher,
Nucl.\ Phys.\ {\bf B400}, 75 (1993)
[hep-ph/9211321].
M.~Ciuchini, E.~Franco, G.~Martinelli and L.~Reina,
Nucl.\ Phys.\ {\bf B415}, 403 (1994)
[hep-ph/9304257].

\bibitem{FSHe} R.~Fleischer,
Z.\ Phys.\ {\bf C58}, 483 (1993).
Z.\ Phys.\ {\bf C62}, 81 (1994).
Z.\ Phys.\ {\bf C66}, 429 (1995)
[hep-ph/9410406].
G.~Kramer, W.~F.~Palmer and H.~Simma,
Nucl.\ Phys.\ {\bf B428}, 77 (1994)
[hep-ph/9402227].
N.~G.~Deshpande and X.~He,
Phys.\ Lett.\ {\bf B336}, 471 (1994)
[hep-ph/9403266].

\bibitem{rpp2000} 
D. E. Groom {\it {et al.}} 
The European Physical Journal {\bf {15}} (2000) 1.

\bibitem{lg} J. Gasser and H. Leutwyler, Phys. Rep. {\bf 87}, 77 (1982).

\bibitem{Atwood:1997bn}
D.~Atwood and A.~Soni,
Phys.\ Lett.\ {\bf B405}, 150 (1997)
[hep-ph/9704357].

\bibitem{Halperin:1998ma}
I.~Halperin and A.~Zhitnitsky,
Phys.\ Rev.\ Lett.\ {\bf 80}, 438 (1998)
[hep-ph/9705251].



\bibitem{Jessop:2000bv}
C.~P.~Jessop {\it et al.}  [CLEO Collaboration],
Phys.\ Rev.\ Lett.\ {\bf 85}, 2881 (2000)
[hep-ex/0006008].


\bibitem{Lipkin:1995hn}
H.~J.~Lipkin,
Phys.\ Lett.\ {\bf B357}, 404 (1995).

\bibitem{Lipkin:1997ke}
H.~J.~Lipkin,
hep-ph/9708253, In Proc. of the 2nd Intern. Conf. on 
B Physics and CP Violation,
Honolulu, Hi, U.S.A., 24-27 mar 1997, editors T. E. Browder, F. A.
Harris and S. Pakvasa, World Scientific, 1998, p.436.

\bibitem{Lipkin:1991us}
H.~J.~Lipkin,
Phys.\ Lett.\ {\bf B254}, 247 (1991).

\bibitem{Lipkin:1992fd}
H.~J.~Lipkin,
Phys.\ Lett.\ {\bf B283}, 421 (1992).

\bibitem{Aleksan:1995bh}
R.~Aleksan, A.~Le Yaouanc, L.~Oliver, O.~Pene and J.~C.~Raynal,
Phys.\ Rev.\ D {\bf 51}, 6235 (1995)
[hep-ph/9408215].


\bibitem{Lipkin:2000sf}
H.~J.~Lipkin,
Phys.\ Lett.\ {\bf B494}, 248 (2000)
[hep-ph/0009241].

\bibitem{Chen:2000hv}
S.~Chen {\it et al.}  [CLEO Collaboration],
Phys.\ Rev.\ Lett.\ {\bf 85}, 525 (2000)
[hep-ex/0001009].

\bibitem{BSW} M. Bauer, B. Stech and M, Wirbel, Z. Phys. {\bf C 34}, 103
(1987).


\bibitem{Datta:1998nr}
A.~Datta, X.~G.~He and S.~Pakvasa,
Phys.\ Lett.\ {\bf B419}, 369 (1998)
[hep-ph/9707259].

\bibitem{Richichi:2000kj}
S.~J.~Richichi {\it et al.}  [CLEO Collaboration],
Phys.\ Rev.\ Lett.\ {\bf 85}, 520 (2000)
[hep-ex/9912059].
\end{thebibliography}
\end{document}